# MODELING HIERARCHICAL STRUCTURES – HIERARCHICAL LINEAR MODELING USING MPLUS

## M. Jelonek

Institute of Sociology, Jagiellonian University
Grodzka 52, 31-044 Kraków, Poland
e-mail: magjelonek@wp.pl

The aim of this paper is to present the technique (and its linkage with physics) of overcoming problems connected to modeling social structures, which are typically hierarchical.
Hierarchical Linear Models provide a conceptual and statistical mechanism for drawing conclusions regarding the influence of phenomena at different levels of analysis.
In the social sciences it is used to analyze many problems such as educational, organizational or market dilemma.
This paper introduces the logic of modeling hierarchical linear equations and estimation based on MPlus software.
I present my own model to illustrate the impact of different factors on school acceptance level.

### 1. General remarks

The aim of my article is not only the description of the hierarchical linear modeling, but also a presentation of common areas of social sciences and physics' interest.
In other words I present what physicists offer (often unconsciously) to social scientists, and what social scientists are searching for (usually apart from physics).
Moreover, I introduce a non-typical mode of multilevel structures modeling by means of software normally used for modeling latent structures – Mplus.
One of the main purposes of this article is to provoke the discussion and an exchange of information between sociology and physics' researchers in the area of hierarchical structures modeling.

## 2. Physics versus Social Sciences – hierarchical modeling

In spite of a long debate about pivotal differences between social and natural sciences and many listed dissimilarities there are a lot of common concerns and dilemmas [1].
For instance, both social scientists and physicists have problem of estimation and modeling, approximation and modeling errors. As J.T. Oden stated: 'The estimation and control of the first of these has been the subject of research for two decades; the estimation and control of the second error source, modeling error, is a relatively new subject, studied in recent years in connection with heterogeneous materials, large deformation of polymers, and inelastic behavior of materials.' [2].
Concerned about relation of social sciences and physics I focused primarily on the question: What physics may offer to the hierarchical structures' modeling in the sociology?
There is some concern about physics regarding multilevel structures [3]:

- We may find some linkage between social sciences and physics, especially in a statistical, computational mechanics ruled by the law of the large numbers but also in a quantum mechanics.
- Physicists focus mostly on the computational part of modeling (algorithms etc).
- They are normally (but there are some exceptions to the rule) interested in the non linear structures modeling [4].
- They are particularly interested in HLM using Bayes' method.
- They designed a software (SAS PROC MIXED, CASPOC) enabling multilevel structures modeling. Above named software wasn't designed specially for social scientists but rather for physicists, biologists, economists, withal SAS PROC MIXED is efficiently exploited within a wide range of social science problems domains.

To sum up, despite the raising interest in sociophysics, a common space for social sciences and physics remains still non explored. I suppose that aforesaid sphere may apply to the hierarchical structures modeling.

## 2. HLM - introduction

Nested, data structures are common throughout many areas of research, not only in social sciences, but also in physics, biology or economy. Especially in sociology hierarchies have a special sense: they structure social life, therefore they reflect an impact of social groups on an individual.

We may find that kind of structure in educational, organizational, family research, cross – national studies, but also in longitudinal, methodological research or growth studies.

Schooling systems present an obvious example of a multilevel framework - students exist within a hierarchical social structure that include family, classroom, school, voivodship, and country.

Students within a particular classroom are similar in values, family background, socio-economic status, or educational preparation. Moreover they share the experience of being in the same environment, which increase their homogeneity.

In most cases researcher is obliged to take into consideration the impact of group on a individual, additionally he is often interested in understanding how group level (environmental) variables (e.g., teaching style, class size, voivodship funding) affect individual outcomes (e.g., achievement, attitudes, etc.).

Until recently these types of data in social sciences were analyzed using typical (improper) for non hierarchical data techniques such as standard regression or structural equation modeling.

With hierarchical linear models, each of the level in a structures is represented by its own submodel expresed how variables at one level influence relations occurring at another.

Most analytical techniques require independence of observations as a primary assumption. Since this assumption is violated in the presence of multilevel data, ordinary least squares regression produces standard errors that are too small. Consequently, this leads to a higher probability of rejection of a null hypothesis.

Moreover, multilevel models are designed to analyze variables from different levels simultaneously, that operation apart from HLM is impossible without simplifications such as data aggregation or disaggregation.

To sum up, hierarchical linear modeling gives us appropriate estimators of level – one and level - two coefficients, corrected standard errors, confidence intervals and statistical tests.

Owing to HLM we may incorporate different levels variables into the common model placing the individual into the group's context.

The goal of this paper is to introduce the problem of methodological approach to the hierarchical structures in social sciences.

I would like to present my own model based on HEALTH BEHAVIOR IN SCHOOL - AGED CHILDREN [5] data to discuss methodological and theoretical problems of taking into account the hierarchy in social structure during the analysis.

I focus on technical aspects of hierarchical linear modeling, therefore such a model serves only as a facilitation of explanation and illustration for crucial problems connected with hierarchical models construction and testing.

## 2. Conceptual model

The aim of my model was to illustrate simultaneously the impact of school and students - level factors on the level of school acceptation.
The dependent variable (school acceptation) was a scale composed of three positions by means of Principal Components Analysis.
Before appropriate model construction I tested hierarchical structure of the data using one – way analysis of variance (ANOVA) and intraclass correlation coefficient.
According to the Muthen's [6] directions I computed the design effect (DEFF) to affirm soundness of multilevel model specification.

DEFF = 1 + (average cluster size – 1) x intraclass correlation

DEFF (3,769) was larger than 2 (minimum for multilevel analysis) therefore I was authorized to build a hierarchical equation.
Additionally I estimated an unconditional model (model without predictors) therefore computing the deviance to help me in the future models evaluation.
Prior to specifying some prediction models, the independent variables were all centered about their respective grand means (centering = grandmean command). That operation allowed the intercepts for the prediction model to be interpreted as the expected average level of school acceptation of an average student at an average high school.
In order to examine the relationship among students and high school characteristics several models was specified and tested.

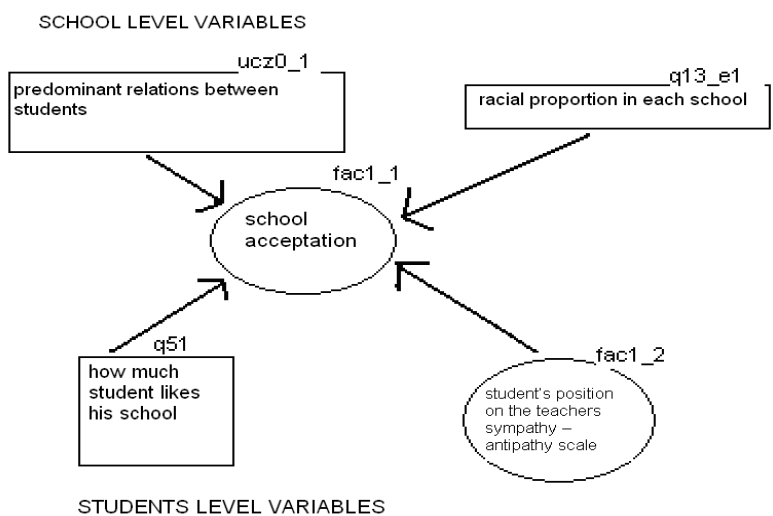

Fig.1. Theoretical two – level model. Impact of school and students - level factors on the level of school acceptation

The first model included only three level - 1 predictors:
- q51 (characterizing how much student likes his school)
- fac1_2 (defining the student's position on the teachers sympathy – antipathy scale)
- q66 (describing how much student is life – satisfied)

Using that random slope model we obtained parameters as with traditional OLS regression, additionally variance in the effect parameters provided an indication of whether there was a sufficient variability in the parameters across high school to build a multilevel model.

The results suggested that q66 variable disturbed the estimation process (even with enlargement of iteration's steps model didn't converge) furthermore wasn't sufficiently correlated with dependent variable. Therefore I decide to remove it from analysis.

By that means, I rebuilt student's level equation (with two predictors), presented below.

$$Fac1\_1 = \beta_{0j} + \beta_{1j} (q51_{ij - mean(q51)}) + \beta_{2j} (fac1\_2_{ij - mean(fac1\_2)}) + r_{ij}$$

The notations 'i' and 'j' signify student 'i' within a 'j' school.

To test the quality of model I used not only prior mentioned deviance - $2log(L_0)$ but also I calculate following Bryk and Raudenbush an index of the proportion reduction in variance [7]

$$R^2 = \sigma^2 (ANOVA) - \sigma^2 (random\ slope\ model)) / \sigma^2 (ANOVA)$$

$$R^2 = 0,924 – 0,575 / 0.924 = 0.38$$

It compare total 'within' variance potentially explainable by any level one model (ANOVA) with a variance explained by a random slope model (potentially explainable after incorporation of level - one predictors into the model).

In our model incorporation of those two variables (q51 and fac1_2) as predictors of school acceptation reduced the within school variance by 38%.

Building the second model - Random Coefficient Regression Model (RCRM) I answered the question: how much do the regression equations vary from school to school (slopes and intercept) that way I tested the randomness of level – one coefficients [8].

The results (extremely small variance - 0.001) suggested the restriction on the q51 coefficient (because of its analytical non randomness).

In the third step I built level - two models for level - one coefficients explanation, by means of two variables.

I identified at least two high school characteristics that may affect the level of school acceptation.

First one, ucz01_1 describing predominant relations between students, typical for each school (0 – school where are dominant negative relations between students, more conflicts than agreements, 1 - school where are dominant positive relations, more agreements than conflicts).

Second, Q13e_1 characterizing the racial proportion in each school (1 – school where are more 'white' than other races students 0 – opposite situation).

Hereby (after models evaluation) I obtained two level - two models with level - two predictors and one non random (level – one) coefficient.

$$\beta_{0j} = \gamma_{00} + \gamma_{01} (ucz0\_1j) + u_{0j}$$
$$\beta_{1j} = \gamma_{10}$$
$$\beta_{2j} = \gamma_{20} + \gamma_{21}(ucz0\_1j) + \gamma_{22}(q13e\_1j) + u_{2j}$$

After incorporation of level – two equations to the level - one model I obtained complete hierarchical model inclosed students level predictors as well as school level ones.

$$Fac1\_1 = \gamma_{00} + \gamma_{01}(ucz0\_1j) + \gamma_{10}(q51_{ij - mean(q51)}) + \gamma_{20}(fac1\_2_{ij - mean(fac1\_2)}) + \gamma_{21}(ucz0\_1j)(fac1\_2_{ij - mean(fac1\_2)}) + \gamma_{22}(q13e\_1j)(fac1\_2_{ij - mean(fac1\_2)}) + u_{0j} + u_{2j}(fac1\_2_{ij - mean(fac1\_2)}) + r_{ij}$$

I decided to estimate parameters by means of restricted maximum likelihood (REML) method.
Because of choice of Maximum Likelihood as the method of estimation, I used the iterative scheme: 'expectation – maximization' (EM) algorithm.
The results are below.

$$Fac1\_1 = 0.046 - 0.090(ucz0\_1j) + 0.382(q51_{ij - mean(q51)}) + 0.328(fac1\_2_{ij - mean(fac1\_2)}) + 0.032(ucz0\_1j)(fac1\_2_{ij - mean(fac1\_2)}) + 0.047(q13e\_1j)(fac1\_2_{ij - mean(fac1\_2)}) + u_{0j} + u_{2j}(fac1\_2_{ij - mean(fac1\_2)}) + r_{ij}$$

1. Interpretation

One of the most complicated operations in hierarchical linear modeling is interpretation of coefficients, especially cross – level terms interpretation. Because of equivocal meaning of those terms researchers typically help out with a theory.
Other coefficients meaning is generally comprehensible and analogous to the interpretation of one – level regression coefficients, particularly when we decided to standardize level – one predictors.

For better understanding, the hierarchical model, equations` construction, imputs` structure and the theory must be pointed out during the interpretation.
Look at an imput of previously discussed model (MPlus software).

MODEL:
    %WITHIN%
    fac11 ON q51;
    s2 |fac11 ON fac12;
    %BETWEEN%
    fac11 s2 ON ucz011 q13e1;

To facilitate the interpretation we may present the same model by means of a simple diagram.

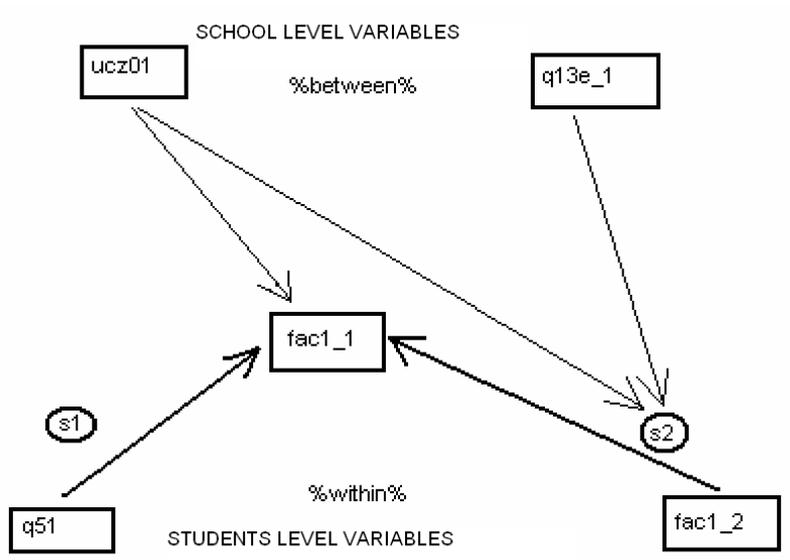

Fig.2. Diagram reflecting the impact of school and students' level factors on the level of school acceptance.

Rectangles represent variables, arrows with circles – the impact of level - one predictors on the dependent variable, circles - level – one coefficients, and other arrows – the impact of level – two predictors on the dependent variable and on the level – one coefficients.
That way I introduced the simplier means of multilevel models presentation, that may be helpful for more elaborate interpretation of hierarchical models.

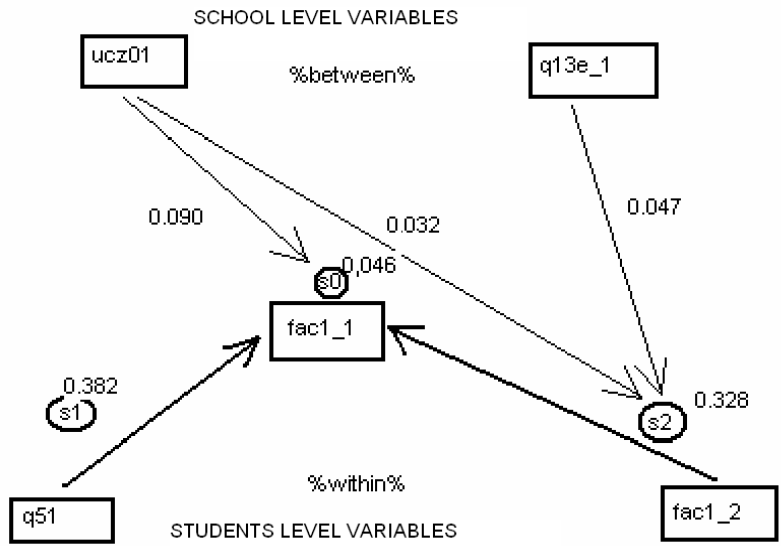

Fig.3. Estimated coefficients for a diagram reflecting the impact of school and students' level factors on the level of school acceptation.

After coefficient estimation, and graphical incorporation to the hierarchical model, interpretation of equations appears to be easier than before.
The most interesting for us will be the sense of cross – level terms.
For instance, reading the model of an impact of school and students - level factors on the level of school acceptation leads us to the conclusions:

- At schools where contacts between students are rather positive than negative I observed stronger correlation between school acceptation and sympathy – antipathy students – teachers relations.
In other words, at schools where there are more conflicts between students, relation: teachers – students exert less influence on school acceptation.
(0,032 as a cross – level term, is the average difference in fac1_2 slope between schools where dominate negative students' relations and schools where dominate positive).

- second cross – level coefficient may be interpreted analogously to the first one. At the school with white race domination there are stronger correlations between school acceptation and sympathy – antipathy students – teachers relations.

## 5. Conclusions

One of the important issues in sociology is the integration of micro and macro concepts, therefore popularization of HLM in Social Sciences should lead to a better understanding of the social structure in all and of its complexity [9].

As the call for developing multilevel theories of social structures continues, it is important to use methodological advances from other disciplines, such as physics.

I suppose that there exists a non explored field in physics useful for social sciences – and that would be hierarchical structures modeling.

Despite the growing interest in modeling hierarchical structures in the whole Europe, HLM in Poland is fairly unknown.

Regardless of patterns of structure, analysis are typically conducted as uni – level. That custom provokes one question: Which of outcomes in sociology is valid? Which of hypothesis should be retested?

**Acknowledgements.** The author thanks Jarosław Górniak for the inspiration and kind help.